\documentclass[12pt]{article}
\usepackage{amsmath,amssymb}

\usepackage{graphicx}
\usepackage{wrapfig}
\usepackage{color}

\usepackage{hyperref}
\usepackage[sort,compress]{cite}

\def\beq{\begin{eqnarray}}
\def\eeq{\end{eqnarray}}

\def\lb{\label}

\newcommand{\Dslash}{D\hskip-0.275truecm/}

\newcommand{\be}{\begin{equation}}
\newcommand{\ee}{\end{equation}}
\newcommand{\bea}{\begin{eqnarray}}
\newcommand{\eea}{\end{eqnarray}}
\newcommand{\bg}{\begin{gather}}

\newcommand{\bseq}{\begin{subequations}}
\newcommand{\eseq}{\end{subequations}}

\renewcommand{\ln}{\mathop{\rm ln}\nolimits}

\def\tr{\hbox{Tr}}

\def\be{\begin{eqnarray}}
\def\ee{\end{eqnarray}}
\def\lb{\label}

\begin{document}

\title{\textbf{
Logarithmic terms in entropy of Schwarzschild black holes in higher 
loops}}
\vspace{1cm}
\author{ \textbf{  Sergey N. Solodukhin}} 

\date{}
\maketitle
\begin{center}
\emph{Institut Denis Poisson, \\
UMR CNRS 7013, Universit\'e de Tours, \\
Parc de Grandmont, 37200 Tours, France}


\end{center}




\vspace{0.2mm}

\begin{abstract}

\noindent { The  Bekenstein-Hawking (BH) entropy is expected to be modified by certain correction terms in the quantum loop expansion.
As is well known  the logarithmic terms in the entropy of black holes appear as a one-loop addition to the classical BH entropy.
In this note we study the further modifications of the logarithmic terms in the entropy of the Schwarzschild black holes due to higher quantum loops:
up to three loops in a general renormalizable theory of gauge fields, scalars and fermions and two loops in quantum gravity.
For a large class of field multiplets (including that of the Standard Model) that include graviton  and for a certain range in the values of the couplings these modifications   manifest themselves  in cooling down the black holes at later stages of evaporation and, respectively, in increasing the life time of the black hole. 
If this picture persists to even higher loops, then the  small black holes formed in the early stages of the cosmic evolution do not evaporate completely by now as is predicted in the standard picture.  Instead, their 
 long-lived (Planckian mass) remnants  are   present in abundance in  today's Universe. 
}

\end{abstract}

\vskip 1 cm
\noindent
\rule{7.7 cm}{.5 pt}\\
\noindent 
\noindent
\noindent ~~~ {\footnotesize e-mails: Sergey.Solodukhin@lmpt.univ-tours.fr}

\pagebreak


Since the works of Bekenstein and Hawking \cite{BH} it is generally accepted that the black holes, considered to be exact solutions to General Relativity,  are characterized by a certain entropy proportional to the area of the horizon. The Schwarzschild black hole is a solution that is described by a single dimensionful parameter, mass $M$ or horizon radius
$r_+=2GM$. Classically, it  has  the entropy
\be
S_{BH}=\frac{A_+}{4G}=4\pi \frac{M^2}{M^2_{PL}}\, ,
\lb{1}
\ee
where $A_+=4\pi r_+^2$ is horizon area and we introduced the Planck mass, $M_{PL}=1/\sqrt{G}$, and the temperature $T_H^{-1}=\partial_M S_{BH}=8\pi M/{M^2_{PL}}$.
This temperature becomes unbounded as soon as mass $M$ of the black hole decreases.  So that the evaporation rate
\be
\frac{dM}{dt}\sim - T^4_H A_+
\lb{1-1}
\ee
accelerates for small black holes as $dM/dt \sim -M^4_{PL}/M^2$ and the black hole evaporates in finite time $t_{BH}\sim M_0^3/M^4_{PL}$, where $M_0$ is the  initial mass.

\bigskip

At the quantum level, when matter  and/or gravity,  are quantized  the classical formula (\ref{1}) gets modified. One way to think about this is to consider the entanglement entropy of the
quantum fields. This entropy requires a UV regulator $\epsilon$ to be properly  defined. The entropy  contains both the area law similar to (\ref{1}) \cite{EQ} and a logarithmic term, which for a Schwarzschild black hole takes a simple form \cite{Solodukhin:1994yz} - \cite{Solodukhin:1997yy},
\be
S_{ent}=\frac{A_+}{48\pi \epsilon^2}+s_0\ln \frac{r_+}{\epsilon}\, ,
\lb{2}
\ee
where the pre-factor in front of  the logarithmic term is
\be
s_0=\frac{1}{45}(N_0+\frac{7}{2}N_{1/2}-13N_1-\frac{233}{4}N_{3/2}+212 N_2+91N_A)
\lb{3}
\ee
for a multiplet of $N_0$ scalars, $N_{1/2}$ Dirac fermions, $N_1$ vector fields, $N_2$ particles of spin 2 (gravitons), $N_{3/2}$ fields of spin $3/2$ (gravitino) and $N_A$ rank 2 antisymmetric tensor fields. 
It should be noted that the existing technique allows us to compute only the UV divergent parts of the entropy. The dependence on $r_+$ in the logarithmic term then comes
from the two facts: that the entropy is a dimensionless quantity  and that there is only one dimensionful parameter, $r_+$ that characterises the geometry. So that the entanglement
entropy may only be a function of  the combination $(r_+/\epsilon)$. The other remark is that  the scalar field contribution in (\ref{3}) to logarithmic term is the same for any value of non-minimal coupling $\xi$ provided the latter is introduced as  $\xi \phi^2 R$.

The quantum entropy (\ref{2}) can be considered as a one-loop correction to the classical entropy (\ref{1}).  In the total entropy, $S_{BH}+S_{ent}$, the UV divergences then can be  absorbed in the renormalization\footnote{There are certain subtleties  \cite{Kabat:1995eq} due to the presence of the non-minimal coupling 
\cite{Nonmin}. These are not, however, essential in the present discussion.
} of
the Newton constant, $1/G_{ren}=1/G+1/12\pi\epsilon^2$,   and the $R^2$ couplings in the gravitational action, see \cite{Ren}.  
So that one finds for the total black hole entropy expressed in terms of the renormalized couplings 
\be
S_{1-loop}=4\pi \frac{M^2}{M^2_{PL}}+s_0\ln M/\mu\, ,
\lb{4}
\ee
where $M_{PL}$ is defined with respect to the renormalized $G_{ren}$ and we omitted the irrelevant constants.
Respectively, the temperature is modified,
\be
T^{-1}=8\pi M/M^2_{PL}+s_0/M\, .
\lb{5}
\ee
The modification becomes important when the mass of the black hole approaches the Planck mass. 
Extrapolating this approximation to those masses one finds that the evaporation scenario  depends on the sign of $s_0$. 

If $s_0=-q<0$ then there exists a certain mass $M_{min}= \sqrt{8\pi q}M_{PL}$ for which the entropy as function of mass
develops a minimum and the temperature becomes infinite (just as for $M=0$ in the classical case). The black hole evaporates down to this minimal mass in a finite time
$t\sim (M_0-M_{min})^5/(M_0^2M^4_{PL})$.  Formally extended to region where $M< M_{min}$ the temperature becomes negative there.

On the other hand, if $s_0>0$ then the entropy is monotonic function of mass while  the temperature develops a maximum at $M_{*}^2= s_0M^2_{PL}/8\pi$, $T_{max}=M_{PL}/\sqrt{32\pi s_0}$. 
Extrapolating the formulas below this mass the temperature decreases and the black hole cools down so that $T\sim M$ for small $M$.
The evaporation rate (\ref{1-1})  then slows down and the black hole evaporates in infinite time. 
Notice that in the case when $s_0\gg 1$ the critical mass $M_*$ can be well above the Planck scale.
For masses $M> M_*$ the specific heat is negative in a similar way as it follows from  the Bekenstein-Hawking  entropy \cite{BH} and, irrespectively of black holes, is in fact valid
for a wide class of gravitationally coupled systems \cite{Thirring}. However, for smaller masses $M< M_*$ the specific heat becomes positive\footnote{In the critical point 
$M=M_*$ the specific heat diverges that makes  it quite attractive to interpret this point as a second order phase transition (for a discussion of singular behavior of specific heat 
in the second order transition see, for instance, \cite{Abe}).  
This is an interesting direction to explore that we leave for a separate study.}.

The entropy (\ref{4}) is not automatically positive function of mass that may signal of the breakdown of the one-loop approximation in the region where the entropy becomes negative.
The entropy is positive (and the approximation is reliable) for $M>\mu$. For sufficiently small $\mu$ (or large values of $s_0$) the threshold point lies well below the critical point $M_*$
so that the most of the interesting modifications discussed above happen well above the point where the approximation breaks down.

\bigskip

The other way to obtain entropy (\ref{4}) is to consider the effective gravitational action,
\be
W_{gr}=-\frac{1}{16\pi G} \int R +W_Q\, 
\lb{5}
\ee
where the quantum part $W_Q$ is represented as an expansion in powers of the Riemann curvature and its covariant derivatives.
It contains both local and non-local terms.  In the one-loop approximation $W_Q=\frac{1}{2}\ln\det {\cal D}$, where ${\cal D}$ is the operator that governs the
quadratic perturbations.
 The entropy then arises as a response of the gravitational action to a small angle deficit $\delta=2\pi (1-\alpha)$
at the horizon $\Sigma$, $S=(\alpha\partial_\alpha-1)W_{gr}|_{\alpha=1}$, see \cite{Solodukhin:2011gn} for more details on this formalism.
The conical singularity manifests itself in the singular terms in the Riemann curvature concentrated at the horizon $\Sigma$, see \cite{Fursaev:1995ef},
\be
R_{\mu\nu\alpha\beta}=2\pi(1-\alpha) ((n_\mu n_\alpha)(n_\nu n_\beta)-(n_\mu n_\beta)(n_\nu n_\alpha)\delta_\Sigma+\dots \, , 
\lb{6}
\ee
where $\dots$ stand for the regular terms in the curvature and $n_\mu^a\, , \, a=1,2$ is a pair of vectors normal to $\Sigma$, $(n_\mu n_\nu)=\sum_{a=1}^2n^a_\mu n^a_\nu$.

Discussing the entropy of the Schwarzschild black holes, for which the Ricci tensor identically vanishes,  we need to look only at  the terms that contain  the Riemann tensor. Focusing on the quadratic terms one finds
\be
W_Q=\frac{s_0}{64\pi^2}\int R_{\mu\nu\alpha\beta} R^{\mu\nu\alpha\beta}\ln\epsilon +\dots\, .
\lb{7}
\ee
Applying formula (\ref{6}) to (\ref{7}) and using that $\int_\Sigma R^{\mu\nu\alpha\beta} n^a_\mu n^b_\nu n^a_\alpha n^b_\beta=8\pi$ one arrives at (\ref{2}).
Notice that $\ln r_+$ term is then restored by the dimensionality arguments. This term, in fact, comes from the non-local (UV finite) part of the quantum action $W_Q$. 

Notice that formula (\ref{4}) is valid for any, not necessarily conformal,  massless field.  However,  if computed in a 4d CFT the logarithmic term takes the form \cite{Solodukhin:2010pk}
\be
s_0=64\pi^2(C-A) \,
\lb{8}
\ee
and is, thus, related to the conformal charges  $A$ and $C$ that appear in the conformal anomaly, $<T>=AE_4-CW^2$, where $E_4$ is the Euler density and $W^2$ is the square of the
Weyl tensor.  Notice that in certain theories $s_0$ vanishes. This is so, for instance, for ${\cal N}=4$ super-Yang-Mills theory.

\bigskip

In the entanglement entropy of massive fields there appear new terms, both UV divergent and finite, that are due to mass $m$. However, most of them go away
in the total entropy expressed in terms of the renormalized Newton constant.  Such terms appear already in Minkowski space-time, see \cite{Kabat:1995eq},
\be
S_{ent}=c_s\frac{A_+}{48\pi}\int_{\epsilon^2} \frac{ds}{s^2}e^{-ms^2}\, ,
\lb{8-1}
\ee
where for a scalar $c_0=1$ and for a Dirac fermion $c_{1/2}=2$. The respective renormalization of the Newton constant goes as
\be
\frac{1}{G_{ren}}=\frac{1}{G}+c_s\frac{A_+}{12\pi}\int_{\epsilon^2} \frac{ds}{s^2}e^{-ms^2}\, .
\lb{8-2}
\ee
In a curved space-time there appear more terms in the entropy that depend on $m$. However, the logarithmic term in the entropy remains unchanged.
This is due to the fact that in the  logarithmically UV divergent terms the  mass $m$ may appear only in a combination $m^2 R$ that contributes to the renormalization of the Newton constant that is already taken into account  in (\ref{8-2}).
There, however, may appear new UV finite terms in the entropy of the form $f(r_+m)$ which grow slower than logarithm\footnote{
For instance, one anticipates an exponential term of the form $e^{-r_+^2m^2}$ which  for a large range of parameters $(M,m)$ leads to smaller effects than the logarithmic term.}.
 Exact form of those terms is  not yet known and is to be determined.
In our analysis we shall ignore these terms concentrating our attention on the logarithmic terms only.

\bigskip

The story described up to this point is what we have at the level of one loop,  when only the interaction of the matter with the background gravitational field  is taken into account.
 The sign of the logarithmic term in (\ref{4}) is essential when we discuss 
the evaporation of the black hole since the logarithmic term becomes important at the later stages of the evaporation when mass of the black hole becomes comparable
to the Planck mass. For negative values of $s_0$ the evaporating black hole reaches, in a finite time, the stage of infinite temperature, just as in the classical case.
The further evolution is difficult to predict within the present analysis.
Since this configuration corresponds to the minimum of the entropy the system  would probably tend to  absorb the energy rather than emit it. Conceptually,
this situation is not an improvement over the classical case.

On the other hand,  in the case of positive $s_0$ the black hole evaporation is much less violent.  
As soon as  the black hole reaches some maximal temperature its mass then further  decreases and the hole starts to quickly cool down. As a result,  in this case, the black holes never evaporate completely, at the later times  they are present as the long-lived  sub-Planckian  remnants. 

 As we see it from (\ref{3}) the sign of $s_0$ is not a priori definite, it depends on the multiplet of fields existing in Nature. Although many fields contribute positively to $s_0$ some of them, as vector fields, contribute negatively. $s_0$ can be computed in the Standard Model which contains 24 fermions, 12 gauge bosons and 4 scalars (a complex doublet). Provided one graviton is added to this multiplet eq. (\ref{3})  gives us
a positive value, $s_{0, SM}=16/5$. Note that without the graviton this value would be negative, $-68/45$. 

The goal of this note is to investigate how the one-loop approximation (\ref{4})  is modified when the interactions are  turned on and the respective higher quantum loops are taken into account.
By   interactions here we mean both the interaction of the matter fields and the purely gravitational interaction when particles of spin 2 are considered. 
In the case of gravity the coupling constant  is dimensionful and we expect there to appear  corrections to (\ref{4}) of the form $M^2_{PL}/M^2$. Identifying those correction terms 
one expands the domain of validity of the approximation.

Our strategy is  first to look at the logarithmic UV divergent terms in the effective action, then compute the respective UV terms in the entropy using (\ref{6}) and finally restore the
dependence on $r_+$ using the dimensionality arguments, in the same line as in one  loop, (\ref{2}) and (\ref{4}).
We base our further analysis  on the  results available in the literature: the instrumental works of Jack and Osborn  \cite{Jack:1983sk} and Jack \cite{Jack:1984sq} -\cite{Jack:1985wd}
(see also a recent paper \cite{Osborn:2016bev} where some typos in the previous publications have been corrected)  for the matter fields and the classical papers by Goroff and Sagnotti and by van de Ven \cite{2-loop} for gravity.  (In the case of gravity the one-loop UV terms were computed in \cite{1-loop} that results in   the graviton contribution in  (\ref{3}).)
We use these earlier results and compute the corresponding modifications in the entropy. Earlier, it was suggested by Sen \cite{Sen:2012dw} that the logarithmic term
is a one-loop effect and it does not get renormalized in higher loops. This is valid for a theory with a dimensionful coupling constant such as quantum gravity.
In the interacting quantum field theories with dimensionless couplings one expects the possible corrections to the one-loop result and, indeed, as we show
below these corrections do appear in the higher loops.


Before proceeding one technical remark is that we use here a regularisation, such as heat kernel or Pauli-Villars,  in which  the UV regulator $\epsilon$ is dimensionful. 
In  the literature one quite often uses the dimensional regularisation, $\epsilon_d=(4-d)$.
The conversion  rule between the two regularisations is the following: $\frac{1}{\epsilon_d}=-\ln \epsilon$.
Earlier works on entanglement entropy in the interacting field systems include \cite{Fursaev:1993qk} - \cite{Akers:2015bgh}.

\bigskip

Below we consider some examples of interacting quantum field theories.

\bigskip

\noindent {\it General renormalizable scalar field theories.} 
Consider a multi-component scalar field $\phi^i\, , \,  i=1, \dots\, , N_0$ with a general Lagrangian
\be
L=\frac{1}{2}(\nabla\phi)^2+V(\phi)\, ,
\lb{10}
\ee
where the potential takes the form
\be
V(\phi)=\frac{1}{4!}\lambda_{ijkl}\phi^i\phi^j\phi^k\phi^l+\frac{1}{3!} g_{ijk}\phi^i\phi^j\phi^k+\frac{1}{2}m_{ij}\phi^i\phi^j+\frac{1}{2}\xi_{ij}\phi^i\phi^j R+\dots \, .
\lb{11}
\ee
The corresponding UV divergences up to 3 loops in this theory have been calculated by Jack and Osborn \cite{Jack:1983sk} in 1984. 
Leaving aside the terms linear in Ricci scalar $R$, that as we discussed above will contribute to the renormalization of  the Newton constant
we are interested in terms which are  quadratic in the Riemann tensor. Those terms in the theory (\ref{10}) appear in one loop, the corresponding contribution to the entropy
is given by the scalar field part in  (\ref{4})  and in three loops\footnote{In two loops the UV term $W^{(2)}_Q$ contains terms at most linear in $R$, $\int g_{ijk}g_{ijk}R$, that contributes to the renormalization of the Newton constant and does not effect the logarithmic term.}, as  was shown by Jack and Osborn\footnote{Notice that our definition for $W_Q$ differs by sign from the one used in  \cite{Jack:1983sk}.},
\be
W^{(3)}_Q=\frac{1}{(4\pi)^6}\frac{\mu^{-\epsilon_d}}{\epsilon_d}\frac{1}{2592}\int \lambda_{ijkl}\lambda_{ijkl} (R_{\alpha\beta\mu\nu} R^{\alpha\beta\mu\nu}-2R_{\mu\nu}R^{\mu\nu}+\frac{1}{3}R^2)+\dots\, ,
\lb{12}
\ee
where we keep only terms quadratic in curvature.  Notice that (\ref{12}) represents the renormalization of the Weyl-square term.
Applying (\ref{6}) we compute the corresponding contribution to the entropy of the Schwarzschild black hole and find that $s_0$ in the three loop approximation
reads
\be
s_0(\lambda)=\frac{N_0}{45}-\frac{1}{(4\pi)^4}\frac{1}{648}\lambda_{ijkl}\lambda_{ijkl}\, .
\lb{13}
\ee
This value of $s_0$ is  smaller than the one for the free scalar fields.

\bigskip

\noindent {\it Gauge theory coupled to fermions and scalars.}  A system of a non-abelian gauge field  $A=A_\mu^a t^a$ coupled to  Dirac fermions $\psi$
and real scalars $\phi$ carrying the representations of a simple gauge group ${\cal G}$ is described by the Lagrangian
\be 
L=\frac{1}{4g^2} \tr (F_{\mu\nu}F^{\mu\nu})+ \frac{1}{2}(D\phi)^T D\phi+\bar{\psi}\Dslash\psi +\frac{1}{2}\bar{\psi}Y_i\psi \phi^i \, ,
\lb{14}
\ee
where $F_{\mu\nu}=\partial_\mu A_\nu-\partial_\nu A_\mu +[A_\mu, A_\nu]$ 
and one included the Yukawa couplings $Y_i$.
 The two loop computation in this theory was performed in \cite{Jack:1984sq}, \cite{Jack:1985wd},  \cite{Jack:1990eb} (some typos have been corrected in \cite{Osborn:2016bev}, see also \cite{Nakayama:2017oye} for further developments and  \cite{Bzowski:2018fql} for a non-perturbative  CFT consideration). As in (\ref{12}) the appropriate UV divergence comes from the renormalization of Weyl-square term,
\be
W^{(2)}_Q=\frac{1}{(4\pi)^4}\frac{\mu^{-\epsilon_d}}{\epsilon_d} \int \left(g^2(\frac{2}{9}C_2^A-\frac{7}{72}C_2^\psi-\frac{1}{18}C_2^\phi)N_1+\frac{1}{96}\tr Y^2\right)R_{\alpha\beta\mu\nu}R^{\alpha\beta\mu\nu}+\dots\, ,
\lb{15}
\ee
where $C_2^A$ ($C_2^\psi$ and $C_2^\phi$) is Casimir operator for gauge fields (respectively for fermions and scalars).
In these notations the beta-function for the gauge coupling, $\beta_g=\frac{g^3}{(4\pi)^2}(-\frac{11}{3} C^A_2 + \frac{2}{3}C_2^\psi + \frac{1}{6} C_2^\phi)$
\cite{tHooft:1998qmr}.  

 So that one finds in the two loop approximation  that
\be
s_0(g, Y)=s_0^{free}-\frac{1}{16\pi^2}\left(g^2(\frac{2}{9}C_2^A-\frac{7}{72}C_2^\psi-\frac{1}{18}C_2^\phi)N_1+\frac{1}{96}\tr Y^2\right)\, .
\lb{16}
\ee
We see that the contributions of the gauge fields and of the Yukawa couplings tend to decrease the value of $s_0$ while
the contributions of the matter fields (fermions and scalars) increase it. In the QCD sector  $C_2^A=3$ and $C_2^\psi=n_F$ (number of flavors)
and no scalars one has that $\frac{2}{9}C_2^A-\frac{7}{72}C_2^\psi=\frac{48-7n_F}{72}$. It is negative if $n_F\geq 7$. In the Standard Model $n_F=6$  and, hence, $s_0(g)<s_0^{free}$.  This observation and that the Yukawa couplings and $\lambda$-couplings in (\ref{13}) make the negative contributions to $s_0$ appear to indicate that in the Standard Model the value of $s_0$ is smaller than in the free theory. To see whether this value remains positive
(provided we included the graviton's contribution in one loop) requires  a more careful analysis including the actual values of all couplings involved\footnote{
In ${\cal N}=4$  $SU(N)$ SYM  the $C$-charge is not expected to renormalise and, hence, the shift in $s_0$ (\ref{16}) is expected to vanish. I thank K. Skenderis for this remark.}.

\bigskip

\noindent {\it Quantum Gravity.}  The one-loop calculation of UV divergences in quantum gravity has been first performed by 't Hooft and Veltman and later confirmed by various authors
\cite{1-loop}. For pure gravity the divergent term in the effective action is
 \be
 W^{(1)}_Q=-\frac{1}{(4\pi)^2}\frac{\mu^{-\epsilon_d}}{\epsilon_d}\int (\frac{53}{45}E_4+\frac{7}{20}R^2_{\mu\nu}+\frac{1}{120}R^2)\, ,
 \lb{17}
 \ee
 where $E_4$ is the Euler density. This term results in graviton's contribution to logarithmic term $s_0$ presented in (\ref{4}).
 
 The two loop computation in quantum gravity was performed by Goroff and Sagnotti and later by van de Ven \cite{2-loop},
 \be
 W^{(2)}_Q=-\frac{209}{2880}\frac{\mu^{-\epsilon_d}}{\epsilon_d}\frac{\kappa}{(4\pi)^4}\int R^{\mu\nu\alpha\beta}R_{\alpha\beta\sigma\rho}R^{\sigma\rho}_{\ \ \ \mu\nu}\, ,
 \lb{18}
 \ee
 where $\kappa=32\pi^2 G$. We keep  here the term that contains only the Riemann tensor. This term is gauge independent and not vanishes on-shell.
 In fact, in two loops there may appear more terms than just (\ref{18}), in particular the products of two Riemann tensors and either Ricci tensor or Ricci  scalar (see paper of Goroff and Sagnotti for some of such terms). The complete information on the exact numerical factors  with which those terms appear in the effective action is not available in the literature.
 These terms, however,  appear to  depend on the gauge and are unlikely to make any contributions to physical quantities such as the entropy.

 Concentrating our attention only on the Riemann tensor we see that the one-loop and two-loop results (\ref{17}) and (\ref{18}) have the same sign. So that the two-loop correction  due to gravitons to $s_0$ comes with the same sign as the one-loop result. Using  (\ref{6}) one finds that on a conical manifold
 \be
 \kappa\int_{{\cal M}_\alpha} R^{\mu\nu\alpha\beta}R_{\alpha\beta\sigma\rho}R^{\sigma\rho}_{\ \ \ \mu\nu}=12\kappa\pi(1-\alpha)\int_\Sigma R_{ab\sigma\rho}R^{\sigma\rho}_{\ \ ab}+\dots=
 192\pi^2(1-\alpha)\frac{\kappa}{r_+^2}+\dots\, ,
 \lb{19}
 \ee
 where $\dots$ stands for terms quadratic in $(1-\alpha)$ and $r_+=2GM$ is the horizon radius. With this relation we find
  \be
 S(M)=4\pi \frac{M^2}{M^2_{PL}}+s(M)\ln\frac{M}{\mu}\, , \ \  s(M)=s_0+\sigma \frac{M^2_{PL}}{M^2}\, , \ \  \sigma=\frac{209}{480}\, 
 \lb{20}
 \ee
  for the entropy of the Schwarzschild black hole
 in the two-loop approximation.
 In agreement with the arguments of  \cite{Sen:2012dw} there is no higher loop gravity corrections to value of $s_0$. However, the two loop computation results in a new $\sigma$-term 
which is negligible for masses  much larger than the Planck mass but it gives an important positive contribution for $M\sim M_{PL}$. This $\sigma$-term, combined with the standard $s_0$-term,  can still be 
regarded as a ``logarithmic term'' keeping in mind that  the coefficient in front of the logarithm now becomes a function of  the mass. In the higher gravitational loops 
there will appear higher powers of $M^2_{PL}/M^2$ in the  function $s(M)$.

 \bigskip  
 
 \noindent{\it Modified temperature and evaporation.}  With the entropy as function of the mass  given by eq.(\ref{20}) one finds the modified temperature
 \be
M_{PL} T^{-1}=\frac{1}{x^{3/2}}(8\pi x^2+s_0 x-\sigma\ln \frac{x}{\lambda})\, , \ \ x=M^2/M_{PL}^2\, , \  \ \lambda=\frac{\mu^2 e}{M^2_{PL}}
\lb{21}
\ee
In the complete analysis  of the temperature  one should study the dependence on three parameters: $s_0$, $\sigma$ and $\lambda$. Here we just mention that
the presence of the $\sigma$-term with $\sigma>0$ makes the temperature positive for any $M$ even if $s_0$ is negative. Indeed, for $-s_0=q>0$ one finds that the function
in the right hand side of (\ref{21}) is everywhere  positive if two conditions are satisfied: 1) $q<16\pi \lambda$ and 2) $\frac{q^2}{32\pi^2\ln{(16\pi\lambda/q)}}<\sigma$. Thus, for sufficiently small values of $q$ the temperature remains  positive function of $M$ and it develops a maximum below which  the evaporating black hole cools down as soon as its mass decreases. Extrapolating (\ref{1-1}) to the process of evaporation with the temperature (\ref{21}) one finds that at later times the mass of the evaporating black hole falls off as $t^{-1/13}$ that is a much slower rate compared to $t^{-1/5}$ in the case of vanishing $\sigma$.
 
 \bigskip

We conclude with some remarks.

\medskip

1.  Strictly speaking, the  extrapolation of the 1- and 2-loop results for the entropy 
to values of mass $M$ approaching the Planck mass
goes beyond the validity of the loop approximation and, thus, is outside the domain of applicability. However, the same and even for more reasons could (and should) be said about the applicability of the classical BH formula for the  entropy. BH formula  is not valid for small black holes of mass $M\sim M_{PL}$ and should be replaced with a better one.
We believe that the higher loop results, presented here and yet approximative,  show us a certain tendency in  the behaviour of $S(M)$  for small $M$ which, likely,  will become even stronger in the higher loops when more  terms in the function $s(M)$ will be available. The higher loop contributions will appear as the higher order  terms in the series with respect to  
$M^2_{PL}/M^2$.
It is possible that, provided  all loops are included, these terms will sum up to something non-analytic at $M=0$.

\medskip

2. The other limitation of our results comes from the fact that here  the quantum corrections to the entropy are computed in a (still classical) Ricci flat geometry.
In a fully consistent consideration one would have to also analyse the modification of the geometry itself due to the quantum corrections. 
On one hand, this is a more difficult problem. On the other, in a non-perturbative analysis of the quantum corrected Einstein equations the black hole may appear to be replaced with a rather different  (and yet  longer-lived) object as the consideration of    \cite{Berthiere:2017tms} shows. These are different indications that  the existence of the long-lived compact objects in the complete theory is a likely possibility.

\medskip

3. The analysis based on the use of the BH entropy plays an important role in the story of primordial black holes (for a review see, for instance, \cite{Carr:2017jsz}). Namely, it imposes a lower limit on the mass of the primordial black holes since the small ones have enough time to evaporate completely until the present days. Provided one uses the modified version of the entropy with the logarithmic terms present 
this conclusion is no more in place: the black hole evaporation slows down as soon as mass decreases below the critical mass $M_*\sim \sqrt{s_0}M_{PL}$ that leads to much longer life time of the black holes.
In this picture the small primordial black holes do not evaporate completely. Instead, all of them  are present in today's Universe as some  long-lived Planckian compact objects. 
From a phenomenological point of view this is a particular case of a more  general conclusion of  \cite{Raidal:2018eoo} about  the much longer life time of black holes for a wide class of possible modifications of the Hawking's formula for the temperature as function of mass $M$.

\section*{Acknowledgements}  I would like to thank B. Pioline for a discussion on the logarithmic terms, K. Krasnov for discussions on the loop calculations in the background field formalism,
and, especially,  M. Shaposhnikov for warm hospitality in his group at  EPFL  and many inspiring discussions during the initial stages of this project. The present work is supported in part by the ERC-AdG-2015 grant 694896.
I also acknowledge the useful communications with I. Jack regarding his papers and with K. Skenderis on renormalization of Weyl-square term in 4d CFTs  and  many insightful conversations with my colleagues  in Tours: M. Chernodub, N. Mohammedi and S. Nicolis.


\begin{thebibliography}{999}

\bibitem{BH} 
  J.~D.~Bekenstein,
  Phys.\ Rev.\ D {\bf 7}, 2333 (1973);
 S.~W.~Hawking,
  Commun.\ Math.\ Phys.\  {\bf 43}, 199 (1975).
  
\bibitem{EQ} 
  G.~'t Hooft,
  Nucl.\ Phys.\ B {\bf 256}, 727 (1985);
 L.~Bombelli, R.~K.~Koul, J.~Lee and R.~D.~Sorkin,
  Phys.\ Rev.\ D {\bf 34}, 373 (1986);
  M.~Srednicki,
  Phys.\ Rev.\ Lett.\  {\bf 71}, 666 (1993).
  C.~G.~Callan, Jr. and F.~Wilczek,
  Phys.\ Lett.\ B {\bf 333}, 55 (1994).
 
  
\bibitem{Solodukhin:1994yz} 
  S.~N.~Solodukhin,
  Phys.\ Rev.\ D {\bf 51}, 609 (1995)
  [hep-th/9407001].
  
\bibitem{Solodukhin:2011gn} 
  S.~N.~Solodukhin,
  Living Rev.\ Rel.\  {\bf 14}, 8 (2011)
  [arXiv:1104.3712 [hep-th]].
  

  
\bibitem{Fursaev:1994te} 
  D.~V.~Fursaev,
  Phys.\ Rev.\ D {\bf 51}, 5352 (1995)
  [hep-th/9412161].
  
\bibitem{Solodukhin:1997yy} 
  S.~N.~Solodukhin,
  Phys.\ Rev.\ D {\bf 57}, 2410 (1998)
  [hep-th/9701106].
  
\bibitem{Sen:2012dw} 
  A.~Sen,
  JHEP {\bf 1304}, 156 (2013)
  [arXiv:1205.0971 [hep-th]].
  
\bibitem{Solodukhin:2010pk} 
  S.~N.~Solodukhin,
  Phys.\ Lett.\ B {\bf 693}, 605 (2010)
  [arXiv:1008.4314 [hep-th]].
  
  
  
\bibitem{Ren} 
L.~Susskind and J.~Uglum,
  Phys.\ Rev.\ D {\bf 50}, 2700 (1994)
  [hep-th/9401070];
  T.~Jacobson,
  gr-qc/9404039;
  D.~V.~Fursaev and S.~N.~Solodukhin,
  Phys.\ Lett.\ B {\bf 365}, 51 (1996)
  [hep-th/9412020];
  J.~G.~Demers, R.~Lafrance and R.~C.~Myers,
  Phys.\ Rev.\ D {\bf 52}, 2245 (1995)
  [gr-qc/9503003];
  
\bibitem{Kabat:1995eq} 
  D.~N.~Kabat,
  Nucl.\ Phys.\ B {\bf 453}, 281 (1995)
  [hep-th/9503016].

  
  \bibitem{Nonmin}
  S.~N.~Solodukhin,
  Phys.\ Rev.\ D {\bf 52}, 7046 (1995)
  [hep-th/9504022];
  F.~Larsen and F.~Wilczek,
  Nucl.\ Phys.\ B {\bf 458}, 249 (1996)
  [hep-th/9506066];
  S.~N.~Solodukhin,
  Phys.\ Rev.\ D {\bf 91}, no. 8, 084028 (2015)
  [arXiv:1502.03758 [hep-th]].
  
  \bibitem{Thirring} W.~Thirring, ``Systems with negative specific heat'', Z.\ Physik. \  {\bf 235}, 339-352 (1970).
  
  \bibitem{Abe}  R. ~Abe, ``Singularity of Specific Heat in the
Second Order Phase Transition'', Progress\  of\ Theoretical \ Physics, {\bf 38}, No. 2, 322-331 (1967).
  
\bibitem{Fursaev:1995ef} 
  D.~V.~Fursaev and S.~N.~Solodukhin,
  Phys.\ Rev.\ D {\bf 52}, 2133 (1995)
  [hep-th/9501127].
  
\bibitem{Fursaev:1993qk} 
  D.~V.~Fursaev,
  Class.\ Quant.\ Grav.\  {\bf 11}, 1431 (1994)
  [hep-th/9309050].
  

  
\bibitem{Hertzberg:2010uv} 
  M.~P.~Hertzberg and F.~Wilczek,
  Phys.\ Rev.\ Lett.\  {\bf 106}, 050404 (2011)
  [arXiv:1007.0993 [hep-th]].
  
\bibitem{Hertzberg:2012mn} 
  M.~P.~Hertzberg,
  J.\ Phys.\ A {\bf 46}, 015402 (2013)
  [arXiv:1209.4646 [hep-th]].
  
\bibitem{Cotler:2015zda} 
  J.~Cotler and M.~T.~Mueller,
  Annals Phys.\  {\bf 365}, 91 (2016)
  [arXiv:1509.05685 [hep-th]];
  J.~S.~Cotler and M.~T.~Mueller,
  arXiv:1512.00023 [hep-th].
  
\bibitem{Akers:2015bgh} 
  C.~Akers, O.~Ben-Ami, V.~Rosenhaus, M.~Smolkin and S.~Yankielowicz,
  JHEP {\bf 1603}, 002 (2016)
  [arXiv:1512.00791 [hep-th]].
  
\bibitem{Jack:1983sk} 
  I.~Jack and H.~Osborn,
  Nucl.\ Phys.\ B {\bf 234}, 331 (1984).

\bibitem{Jack:1984sq} 
  I.~Jack,
  Nucl.\ Phys.\ B {\bf 234}, 365 (1984).
  
\bibitem{Jack:1985wd} 
  I.~Jack,
  Nucl.\ Phys.\ B {\bf 253}, 323 (1985).
  
\bibitem{Jack:1990eb} 
  I.~Jack and H.~Osborn,
  Nucl.\ Phys.\ B {\bf 343}, 647 (1990).

\bibitem{Osborn:2016bev}
  H.~Osborn and A.~Stergiou,
  JHEP {\bf 1606} (2016) 079
  [arXiv:1603.07307 [hep-th]].
  
\bibitem{Nakayama:2017oye} 
  Y.~Nakayama,
  JHEP {\bf 1707}, 004 (2017)
  [arXiv:1702.02324 [hep-th]].
  
\bibitem{Bzowski:2018fql} 
  A.~Bzowski, P.~McFadden and K.~Skenderis,
  JHEP {\bf 1811}, 159 (2018)
  [arXiv:1805.12100 [hep-th]].

\bibitem{tHooft:1998qmr}
G.~'t Hooft,
  Nucl.\ Phys.\ Proc.\ Suppl.\  {\bf 74}, 413 (1999)
  [hep-th/9808154].
  
\bibitem{1-loop} 
  G.~'t Hooft and M.~J.~G.~Veltman,
  Ann.\ Inst.\ H.\ Poincare Phys.\ Theor.\ A {\bf 20}, 69 (1974);
  S.~M.~Christensen and M.~J.~Duff,
  Nucl.\ Phys.\ B {\bf 170}, 480 (1980);
  G.~W.~Gibbons and M.~J.~Perry,
  Nucl.\ Phys.\ B {\bf 146}, 90 (1978);
  E.~S.~Fradkin and A.~A.~Tseytlin,
  Nucl.\ Phys.\ B {\bf 227}, 252 (1983).
  
  
\bibitem{2-loop}
 M.~H.~Goroff and A.~Sagnotti,
  Nucl.\ Phys.\ B {\bf 266}, 709 (1986);
  A.~E.~M.~van de Ven,
  Nucl.\ Phys.\ B {\bf 378}, 309 (1992).
  
\bibitem{Berthiere:2017tms} 
  C.~Berthiere, D.~Sarkar and S.~N.~Solodukhin,
  Phys.\ Lett.\ B {\bf 786}, 21 (2018)
  [arXiv:1712.09914 [hep-th]].
  
\bibitem{Carr:2017jsz} 
  B.~Carr, M.~Raidal, T.~Tenkanen, V.~Vaskonen and H.~Veermдe,
  Phys.\ Rev.\ D {\bf 96}, no. 2, 023514 (2017)
  [arXiv:1705.05567 [astro-ph.CO]].
  
\bibitem{Raidal:2018eoo} 
  M.~Raidal, S.~Solodukhin, V.~Vaskonen and H.~Veermдe,
  Phys.\ Rev.\ D {\bf 97}, no. 12, 123520 (2018)
  [arXiv:1802.07728 [astro-ph.CO]].
  

\end{thebibliography}
\end{document}